\documentclass{ckm}                 

\confname{Workshop on the CKM Unitarity Triangle, IPPP Durham, April
  2003}

\title{Is the Unitarity of the quark-mixing-CKM-matrix violated in neutron $\beta$-decay? }

 \author{H. Abele}
 \address{Physikalisches Institut der
Universit\"at Heidelberg, Philosophenweg 12, 69120 Heidelberg,
Germany}

\begin{document}

\begin{abstract}
Measurements by various international groups of researchers
determine the strength of the weak interaction of the neutron,
which gives us unique information on the question of the quark
mixing. Neutron $\beta$-decay experiments now challenge the
Standard Model of elementary particle physics with a deviation,
2.7 times the stated error.
\end{abstract}

\maketitle


\section{The Standard Model, quark mixing and the CKM matrix}
This article is about the interplay between the Standard Model of
elementary particle physics and neutron $\beta$-decay. Since the
Fermi decay constant is known from muon decay, the Standard Model
describes neutron $\beta$-decay with only two additional
parameters. One parameter is the first entry $|V_{ud}|$ of the
CKM-matrix. The other one is $\lambda$, the ratio of the vector
coupling constant and the axial vector constant. In principle, the
ratio $\lambda$ can be determined from QCD lattice gauge theory
calculation, but the results of the best calculations vary by up
to 30\%. In neutron decay, several observables are accessible to
experiment, which depend on these parameters, so the problem is
overdetermined and, together with other data from particle and
nuclear physics, many tests of the Standard Model become possible.
$|V_{ud}|$ results significantly from the neutron lifetime $\tau$
and the $\beta$-asymmetry parameter $A_0$.

As is well known, the quark eigenstates of the weak interaction do
not correspond to the quark mass eigenstates. The weak eigenstates
are related to the mass eigenstates in terms of a 3 x 3 unitary
matrix $V$, the so called Cabibbo-Kobayashi-Maskawa (CKM) matrix.
By convention, the u, c and t quarks are unmixed and all mixing is
expressed via the CKM matrix ${\it V}$ operating on d, s and b
quarks. The values of individual matrix elements are determined
from weak decays of the relevant quarks. Unitarity requires that
the sum of the squares of the matrix elements for each row and
column be unity. So far precision tests of unitarity have been
possible for the first row of ${\it V}$, namely
\begin{equation} |V_{ud}|^2 + |V_{us}|^2 + |V_{ub}|^2 = 1-\Delta
\end{equation} In the Standard Model, the CKM matrix is unitary
with $\Delta$ = 0.

A violation of unitarity in the first row of the CKM matrix is a
challenge to the three generation Standard Model. The data
available so far do not preclude there being more than three
generations; CKM matrix entries deduced from unitarity might be
altered when the CKM matrix is expanded to accommodate more
generations \cite{Groom,Marciano1}. A deviation $\Delta$ has been
related to concepts beyond the Standard Model, such as couplings
to exotic fermions \cite{Langacker1,Maalampi}, to the existence of
an additional Z boson \cite{Langacker2,Marciano2} or to the
existence of right-handed currents in the weak interaction
\cite{Deutsch}. A non-unitarity of the CKM matrix in models with
an extended quark sector give rise to an induced neutron electric
dipole moment that can be within reach of next generation of
experiments \cite{Liao}.

Due to its large size, a determination of $|V_{ud}|$ is most
important. It has been derived from a series of experiments on
superallowed nuclear $\beta$-decay through determination of phase
space and measurements of partial lifetimes. With the inclusion of
nuclear structure effect corretions a value of $|V_{ud}|$ =
0.9740(5)~\cite{Hardy} emerges in good agreement of different,
independent measurements in nine nuclei. Combined with $|V_{us}|$
= 0.2196(23) from kaon-decays and $|V_{ub}|$ = 0.0036(9) from
B-decays, this lead to $\Delta$ = 0.0032(14), signaling a
deviation from the Unitarity condition by 2.3 $\sigma$ standard
deviation. The quoted uncertainty in $|V_{ud}|$, however, is
dominated by theory due to amount, size and complexity of
theoretical uncertainties. Although the radiative corretions
include effects of order Z$\alpha^2$, part of the nuclear
corrections are difficult to calculate. Further, the change in
charge-symmetry-violation for quarks inside nuclei results in an
additional change in the predicted decay rate which might lead to
a systematic underestimate of $|V_{ud}|$. A limit has been reached
where new concepts are needed to progress. Such are offered by
studies with neutron and with limitations with pion $\beta$-decay.
The pion $\beta$-decay has been measured recently at the PSI. The
pion has a different hadron structure compared with neutron or
nucleons and it offers an other possibility in determining
$|V_{ud}|$. The preliminary result is
$|V_{ud}|$=0.9971(51)~\cite{Pocanic}. The somewhat large error is
due to the small branching ratio of 10$^{-8}$.

Further information on the CKM matrix and the unitarity triangle
are based on a workshop held at CERN \cite{Battaglia} and a
workshop held at Heidelberg \cite{HADM}.
\section{Neutron-$\beta$-decay}
In this article, we derive $|V_{ud}|$, not from nuclear
$\beta$-decay, but from neutron $\beta$-decay data. In this way,
the unitarity check of (1) is based solely on particle data, i.e.
neutron $\beta$-decay, K-decays, and B-decays, where theoretical
uncertainties are significantly smaller. So much progress has been
made using highly polarized cold neutron beams with an improved
detector setup that we are now capable of competing with nuclear
$\beta$-decays in extracting a value for $V_{ud}$, whilst avoiding
the problems linked to nuclear structure.
\begin{figure}
\hbox to\hsize{\hss
\includegraphics[width=\hsize]{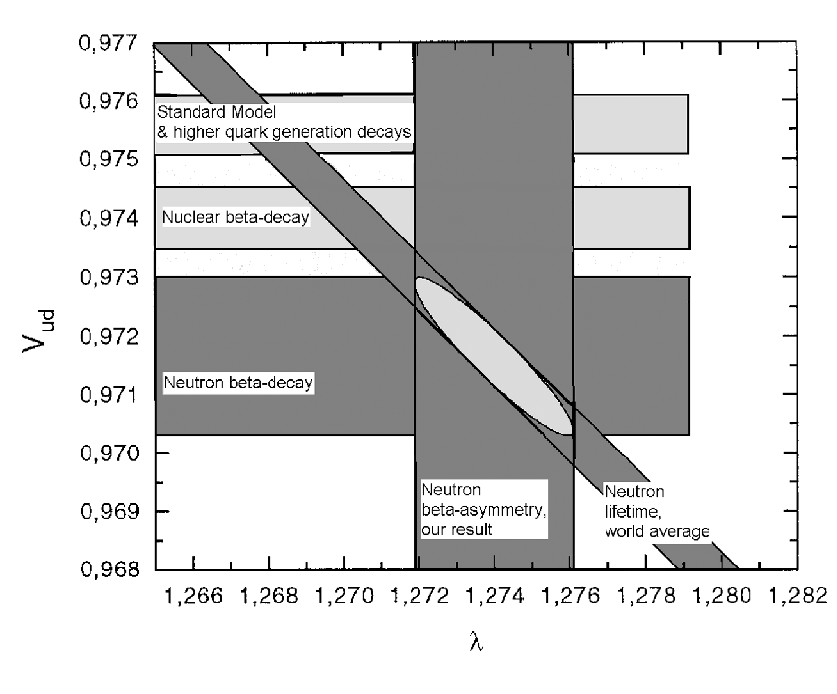}
\hss} \caption{$|V_{ud}|$  vs. $\lambda$. $|V_{ud}|$ was derived
from $Ft$ values of nuclear $\beta$-decays, higher quark
generation decays, assuming the unitarity of the CKM matrix, and
neutron $\beta$-decay.}
\end{figure}
A neutron decays into a proton, an electron and an electron
antineutrino. Observables are the neutron lifetime $\tau$ and
spins $\sigma_e$, $\sigma_\nu$, $\sigma_p$, and momenta $p_e$,
$p_\nu$, $p_p$ of the electron, antineutrino and proton
respectively. The electron spin, the proton spin and the
antineutrino are not usually observed. The lifetime is given by
\begin{equation}
\tau^{-1}=C |V_{ud}|^2(1+3\lambda^2) f^R(1+\Delta_R),
\end{equation}
where $C=G_F^2 m_e^5/(2 \pi^3)=1.1613\cdot10^{-4} s^{-1}$ in
$\hbar=c=1$ units, $f^R$ =1.71335(15) is the phase space factor
(including the model independent radiative correction) adjusted
for the current value of the neutron-proton transition energy and
corrected by Marciano \cite{Marciano}. $\Delta_R$ = 0.0240(8) is
the model dependent radiative correction to the neutron decay rate
\cite{Towner1}. The $\beta$-asymmetry $A_0$ is linked to the
probability that an electron is emitted with angle $\vartheta$
with respect to the neutron spin polarization $P$ = $<\sigma_z>$:
\begin{equation} W(\vartheta) = 1 +\frac{v}{c}PA\cos(\vartheta),
\end{equation} where $v/c$ is the electron velocity expressed in
fractions of the speed of light. ${\it A}$ is the
$\beta$-asymmetry coefficient which depends on $\lambda$. On
account of order 1\% corrections for weak magnetism, $g_V-g_A$
interference,
and nucleon recoil, ${\it A}$ has the form $A$ = $A_0$(1+$A_{{\mu}m}$($A_1W_0+A_2W+A_3/W$)) 
with electron total energy $W = E_e /m_ec^2+1$ (endpoint $W_0$).
$A_0$ is a function of $\lambda$
\begin{equation}  A_0=-2\frac{\lambda(\lambda+1)}{1+3\lambda^2},
\end{equation} where we have assumed that $\lambda$ is real. The coefficients $A_{{\mu}m}$, $A_1$, $A_2$, $A_3$ are from \cite{Wilkinson1} taking a different $\lambda$ convention into consideration. In addition, a further small radiative
correction \cite{Gluck} of order 0.1\% must be applied. For
comparison, information about $|V_{ud}|$ and $\lambda$ are shown
in Fig. (1). The bands represent the one sigma error of the
measurements. The $\beta$-aymmetry $A_0$ in neutron decay depends
only on $\lambda$, while the neutron lifetime $\tau$ depends both
on $\lambda$ and $|V_{ud}|$. The intersection between the curve,
derived from $\tau$ and $A_0$, defines $|V_{ud}|$ within one
standard deviation, which is indicated by the error ellipse. Other
information on $|V_{ud}|$, derived from nuclear $\beta$-decay and
higher quark generation decays, assuming the unitarity of the CKM
matrix, are shown, too. As can be seen from Fig. (1), both the
nuclear $\beta$-decay result from~\cite{Hardy} and the neutron
$\beta$-decay from~\cite{Abele1} do not agree with this unitarity
value.

\section{The experiment PERKEO and the result for $|V_{ud}|$}

The following section is about our measurement of the neutron
$\beta$-asymmetry coefficient $A$ with the instrument PERKEOII,
and on the consequences for $|V_{ud}|$. The strategy of PERKEOII
followed the instrument PERKEO~\cite{Bopp} in minimizing
background and maximizing signal with a $4\pi$ solid angle
acceptance over a large region of the beam. Major achievements of
the instrument PERKEO are:
\begin{itemize}
\item The signal to background ratio in the range of interest is
200. \item The overall correction of the raw data is 2.04\%. \item
The detector design allows an energy calibration with linearity
better than 1\%. \item New polarizers and developments in
polarization analysis led to smaller uncertainties related to
neutron beam polarization.
\end{itemize}
\begin{figure}
\hbox to\hsize{\hss
\includegraphics[width=\hsize]{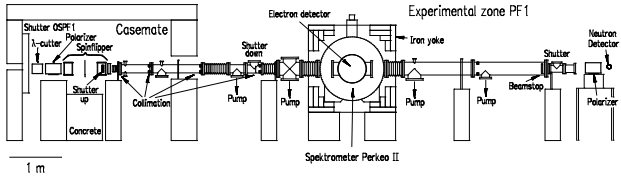}
\hss} \caption{A schematic view of the whole setup at the Institut
Laue-Langevin in Grenoble.} \label{setup}
\end{figure}

For a measurement of $\beta$-asymmetry $A_0$, the instrument
PERKEO was installed at the PF1 cold neutron beam position at the
High Flux Reactor at the Institut Laue-Langevin, Grenoble. Cold
neutrons are obtained from a 25 K deuterium cold moderator near
the core of the 57 MW uranium reactor. The neutrons are guided via
a 60 m long neutron guide of cross section 6 x 12 cm$^2$ to the
experiment and are polarized by a 3 x 4.5 cm$^2$ supermirror
polarizer. The de Broglie wavelength spectrum of the cold neutron
beam ranges from about 0.2 nm to 1.3 nm. The degree of neutron
polarization was measured to be P = 98.9(3)\% over the full cross
section of the beam. The polarization efficiency remained constant
during the whole experiment. The neutron polarization is reversed
periodically with a current sheet spin flipper. The main component
of the PERKEO II spectrometer is a superconducting 1.1 T magnet in
a split pair configuration, with a coil diameter of about one
meter. Neutrons pass through the spectrometer, whereas decay
electrons are guided by the magnetic field to either one of two
scintillation detectors with photomultiplier readout. The detector
solid angle of acceptance is truly 2x2$\pi$ above a threshold of
60 keV. Electron backscattering effects, serious sources of
systematic error in $\beta$-spectroscopy, are effectively
suppressed. Technical details about the instrument can be found in
\cite{Abele,Reich}. The measured electron spectra
$N^\uparrow_i(E_e)$ and $N^\downarrow_i(E_e)$ in the two detectors
(i=1,2) for neutron spin up and down, respectively, define the
experimental asymmetry as a function of electron kinetic energy
$E_e$ and are shown in Fig. 3. \begin{equation}
A_{i_{exp}}(E_e)=\frac{N^\uparrow_i(E_e) -
N^\downarrow_i(E_e)}{N^\uparrow_i(E_e) + N^\downarrow_i(E_e)}.
\end{equation}
By using (3) and with $<\cos(\vartheta)>$ = 1/2, $A_i{_{exp}}(E)$
is directly related to the asymmetry parameter
\begin{equation}
A_{exp}(E_e)=A_{1_{exp}}(E_e)-A_{2_{exp}}(E_e)=\frac{v}{c}APf.
\end{equation}
\begin{figure}
\hbox to\hsize{\hss
\includegraphics[width=\hsize]{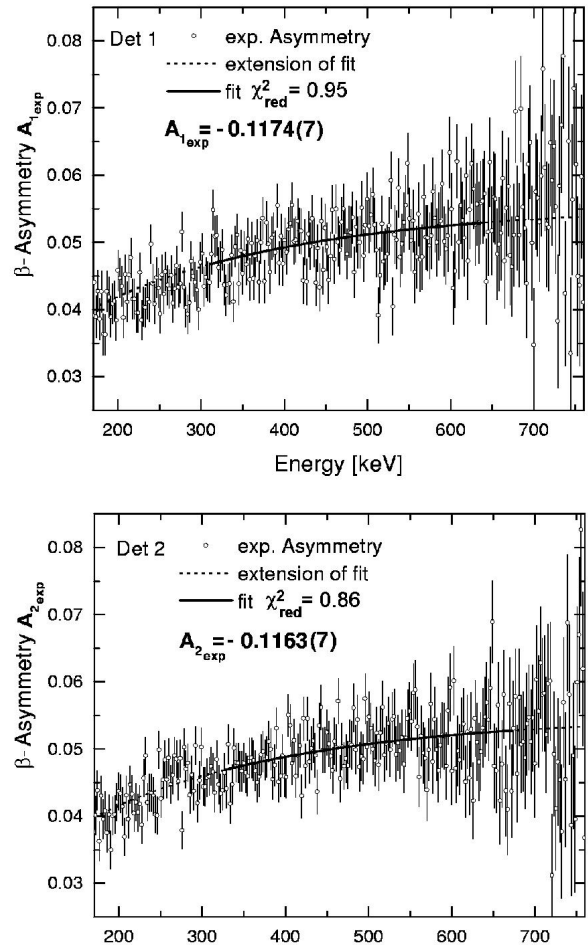}
\hss} \caption{Fit to the experimental asymmetry $A_{exp}$ for
detector 1 and detector 2. The solid line shows the fit interval,
whereas the dotted line shows an extrapolation to higher and lower
energies.}\label{fit}
\end{figure}
The experimental function $A_{i_{exp}}(E_e)$ and a fit with one
free parameter $A_{i_{exp}}$ (the absolute scale of $A_0$) is
shown in Fig. \ref{fit}. The total correction for the small
experimental systematic effects is 2.04\%.

With recent experiments from the University of Heidelberg
\cite{Abele1,Abele}, we obtain $A_0$ = -0.1189(7) and $\lambda$ =
-1.2739(19). With this value, and the world average for $\tau$ =
885.7(7) s, we find that $|V_{ud}|$ = 0.9724(13). With $|V_{us}|$
= 0.2196(23) and the negligibly small $|V_{ub}|$ = 0.0036(9), one
gets
\begin{equation} |V_{ud}|^2 +
|V_{us}|^2 + |V_{ub}|^2 = 1 - \Delta = 0.9924(28).
\end{equation} This value differs from the Standard
Model prediction by $\Delta$ = 0.0076(28), or 2.7 times the stated
error. Earlier experiments \cite{Bopp,Yerozolimsky,Schreckenbach}
gave significant lower values for $\lambda$. Averaging over our
new result and previous results, the Particle Data Group
\cite{Groom} arrives at a new world average for $|V_{ud}|$ from
neutron $\beta$-decay which leads to a 2.2 s deviation from
unitarity.

An independent test of CKM unitarity comes from W physics at LEP
\cite{Sbarra} where W decay hadronic branching ratios can be used
expressed in terms of
\begin{equation}
\frac{Br(W\rightarrow q\bar{q})}{1-Br(W\rightarrow
\bar{q})}=(1+\frac{\alpha}{\pi}\sum{|V_{ij}|^2}).
\end{equation}
Since decay into the top quark channel is forbidden by energy
conservation one would expect $\sum{|V_{ij}|^2}$ to be 2 with a
three generation unitary CKM matrix. The experimental result is
2.039(25), consistent with (7) but with considerably lower
accuracy.
\section{The future}
The main corrections in the experiment PERKEO are due to neutron
beam polarization (1.1\%), background (0.5\%) and flipper
efficiency (0.3\%). The total correction is 2.04\%. With such
small corrections to the data, we start to see a deviation from
the Standard Model already in the uncorrected raw data. For the
future, the plan is further to reduce all corrections. In the
meantime, major improvements both in neutron flux and degree of
neutron polarization has been made: First, the new ballistic
supermirror guide at the ILL from the University of Heidelberg
gives an increase of a factor of 6 in the cold neutron
flux~\cite{Haese}. Second, a new arrangement of two supermirror
polarizers allows to achieve an unprecedented degree of neutron
polarization $P$ of between 99.5\% and 100\% over the full cross
section of the beam~\cite{Soldner}. Third, systematic limitations
of polarization measurements have been investigated: The beam
polarization can now be measured with a completely new method
using an opaque $^3$He spin filter with an uncertainty of 0.1\%
\cite{Heil,Zimmer}. As a consequence, we are now in the lucky
situation to improve on the main uncertainties in reducing the
main correction of 1.1ß\% to less than 0.5\% with an error of
0.1\%. Thus, a possible deviation from the Standard Model, if
confirmed, will be seen very pronounced in the uncorrected data.
Future trends have been presented on the workshop "Quark-mixing,
CKM Unitarity" in Heidelberg from 19 to 20 September 2002.
Regarding the Unitarity problem, about half a dozen new
instruments are planed or are under construction to allow for
beta-neutrino correlation $a$ and beta-correlation $A$
measurements at the sub-10$^{-3}$ level.
\section{Summary}
$|V_{ud}|$, the first element of the CKM matrix, has been derived
from neutron decay experiments in such a way that a unitarity test
of the CKM matrix can be performed based solely on particle
physics data. With this value, we find a 2.7 $\sigma$ standard
deviation from unitarity, which conflicts the prediction of the
Standard Model of particle physics.

Future trends have been presented on the workshop "Quark-mixing,
CKM Unitarity" in Heidelberg, September 19-20, 2002. Regarding the
Unitarity problem, about half a dozen new instruments are planed
or under construction to allow for beta-neutrino correlation $a$
and $\beta$-asymmetry $A$ measurements at the sub-10$^{-3}$ level.
With next generation experiments measurements with a decay rate of
1 MHz become feasible \cite{Dubbers3}.

This work has been funded in part by the German Federal Ministry
(BMBF) under contract number 06 HD 153 I.

\end{document}